\documentstyle[aps,prl]{revtex}
\begin{document}

\twocolumn[\hsize\textwidth\columnwidth\hsize\csname
@twocolumnfalse\endcsname

\title{The formation, ripening and stability of epitaxially strained island arrays}
\author{Helen R. Eisenberg\cite{email1} and Daniel Kandel\cite{email2}  }
\address{Department of Physics of Complex Systems,\\
Weizmann Institute of Science, Rehovot 76100, Israel}

\maketitle

\begin{abstract}
We study the formation and evolution of coherent islands on lattice
mismatched epitaxially strained films. Faceted islands form in films
with aniostropic surface tension. Under annealing, these islands ripen
until a stable array is formed, with an island density which increases
with film thickness. Under deposition, an island shape transition
occurs, which leads to a bimodal island size distribution. In films
with isotropic surface tension we observe continual ripening of islands
above a certain film thickness. A stable wavy morphology is found in
thinner films.
\newline

\end{abstract}

\newline
]
\input epsf
Coherent (dislocation-free) islands form to relieve the strain
associated with lattice mismatched heteroepitaxial thin films.
Their subsequent self-assembly into periodic arrays is of great
interest as the arrays can be used to create quantum dot
structures of importance in semiconductor and optoelectronic
devices. Such island arrays must have a narrow size distribution
in order to be of use in applications. Of particular interest is
whether the island arrays that form are energetically stable or
metastable configurations that will ripen. Here we show in
annealing simulations, that anisotropy in surface tension is
necessary for the formation of a stable (roughly periodic) array
with a narrow size distribution. Moreover, we show that the
presence of a cusp in the surface energy is essential for
reproducing the experimentally observed increase in island density
with increasing film thickness. We also show that a single cusp in
the surface energy (along with elastic relaxation) is sufficient
in order to explain the island shape transition \cite{ross}, which
occurs in growth experiments, and the associated bimodal island
size distribution.

We study the evolution of an elastically isotropic system using
continuum theory. The surface of the solid is at $y=h(x,t)$ and the
film is in the $y>0$ region with the film-substrate interface at $y=0$.
The system is modeled to be invariant in the $z$-direction, and all
quantities are calculated for a section of unit width in that
direction. This is consistent with plane strain where the solid extends
infinitely in the $z$-direction and hence all strains in this direction
vanish.

We assume that surface diffusion is the dominant mass transport
mechanism, leading to the following evolution equation \cite{mullins}:
\begin{equation}
\frac{\partial h}{\partial t}=\frac{D_{s}\eta \Omega}{k_{B}T}
\frac{\partial }{\partial x}\frac{\partial \mu }{\partial s},
\label{evol}
\end{equation}
where $D_{s}$ is the surface diffusion coefficient, $\eta $ is the
number of atoms per unit area on the solid surface, $\Omega$ is the
atomic volume, $T$ is the temperature, $k_{B}$ is the Boltzmann
constant, $s$ is the arc length and $\mu$ is the chemical potential at
the surface.

In our previous work \cite{eisenberg1,eisenberg2} we showed that $\mu$
can be expressed as
\begin{eqnarray}
\frac{\mu}{\Omega}&=&\widetilde{\gamma }(\theta)\kappa +\frac{\
df_{el}^{(0)}}{dh} \nonumber \\
&\;&+\left. \left( \frac{1}{2}S_{ijkl}\sigma _{ij}\sigma
_{kl}-\frac{1}{2} S_{ijkl}\sigma _{ij}^{(0)}\sigma _{kl}^{(0)}\right)
\right| _{y=h(x)}, \label{dfdh3}
\end{eqnarray}
where $\kappa $ is surface curvature, $\theta$ is the angle between the
normal to the surface and the $y$-direction and
$\widetilde{\gamma}(\theta )=\gamma (\theta)+\partial^{2}\gamma
/\partial \theta^{2}$ is the surface stiffness (with $\gamma(\theta)$
being the surface tension). $S_{ijkl}$ are the compliance coefficients
of the material, $\sigma_{ij}$ is the total stress in the material,
$\sigma_{ij}^{(0)}$ is the mismatch stress in the zero strain reference
state and $f_{el}^{(0)}(h)$ is the reference state free energy per unit
length in the $x$-direction. The reference state is defined as a flat
film of thickness $h$ confined to have the lateral lattice constants of
the substrate.

Linear stability analysis predicts that a flat film thinner than the
linear wetting layer thickness, $h_c$, is stable at all perturbation
wavelengths and is marginally stable to perturbations of wavelength
$\lambda_c$ for thickness $h_c$. The expressions for $h_c$ and
$\lambda_c$ are given in \cite{eisenberg1,eisenberg2}. Above $h_{c}$
the flat film is unstable to a larger and larger range of wavelengths
$\lambda_{-}\leq \lambda \leq \lambda_{+}$ until for infinitely thick
films the film is unstable to all perturbations of wavelengths larger
than $\lambda=\lambda_c/2$.

We simulated the surface evolution given by Eqs.\ (\ref{evol}) and
(\ref{dfdh3}) using the numerical scheme described in our earlier work
\cite{eisenberg2}. We used the cusped form of surface tension given by
Bonzel and Preuss \cite{bonzel}, which shows faceting in a free crystal
at 0$^{\circ},\pm 45^{\circ }$ and $\pm 90^{\circ }$.
$df_{el}^{(0)}(h)/dh$ was obtained from ab-initio quantum mechanical
calculations of Si$_{1-x}$Ge$_{x}$ grown on Si(001) (for details see
\cite{eisenberg3}). All our simulations start from a randomly perturbed
flat film with an initial thickness denoted by $C$.

\begin{figure}[h]
\centerline{
\epsfxsize=90mm
\epsffile{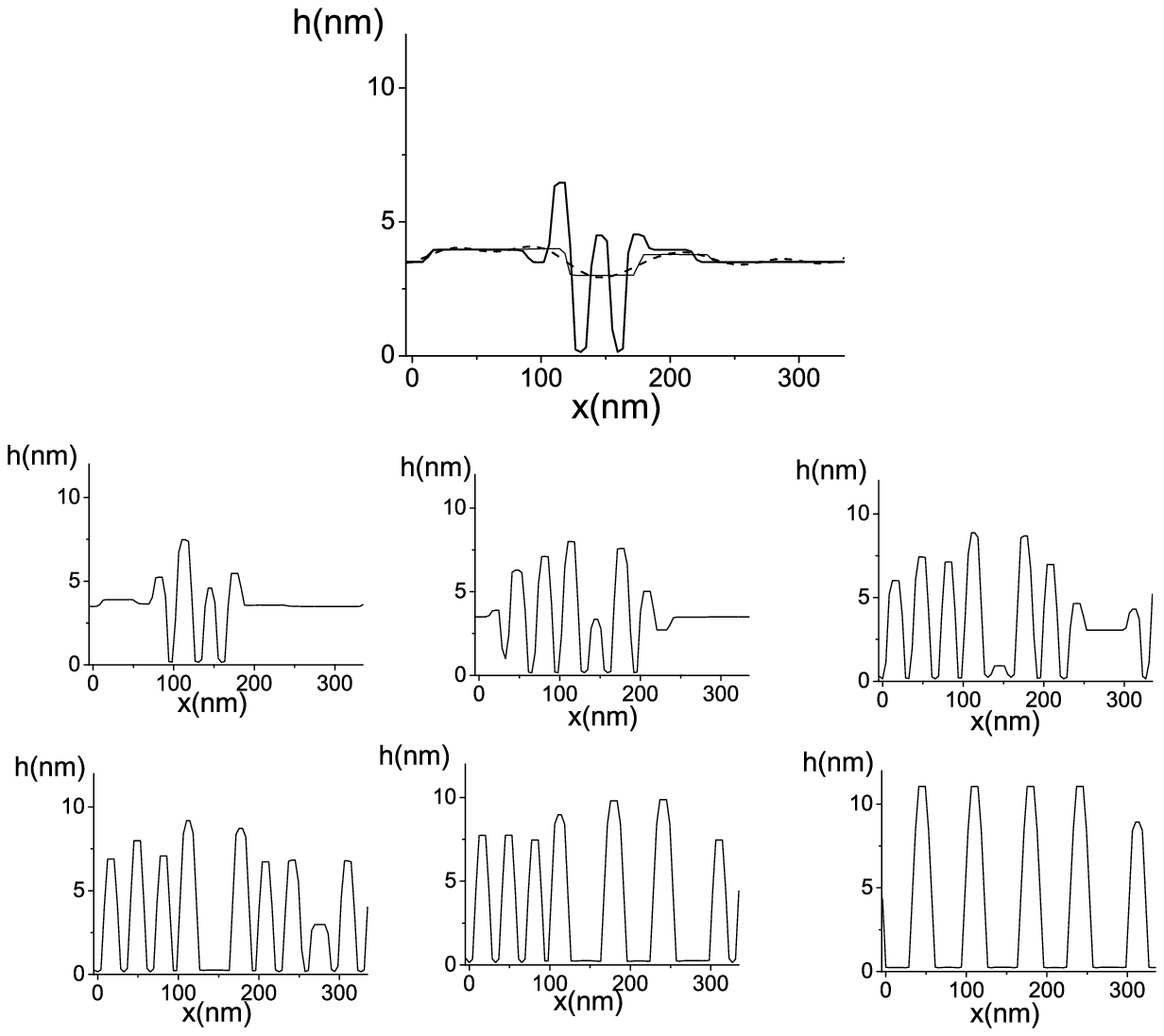}}
\caption{Evolution of
a random perturbation on a 25 monolayer thick Ge film on a Si(001)
substrate. In the first graph the dashed line is the initial
perturbation, the thin solid line is the surface at t=0.005s and the
thick solid line is the surface at t=0.032s. The second to sixth graphs
show the surface at times t=0.044s, 0.068s, 0.094s, 0.123s and 2.181s.
The final graph is the stable steady state island array. Note the
ripple effect in island formation and the later island ripening leading
to a stable island array.}
     \label{facevolc25}
\end{figure}

When perturbations larger than a critical amplitude
\cite{eisenberg1,eisenberg2,eisenberg3} are applied to a flat film,
faceted islands develop in the film during both annealing and growth,
as illustrated in Fig.\ \ref{facevolc25}. The film first becomes
unstable at wavelength $\lambda\sim 50 \gamma(0)/M\varepsilon^2$, where
$M$ is the plain strain modulus and $\varepsilon$ is the lattice
mismatch. The islands which form from this perturbation typically have
a width of about $10 \%$ of the unstable wavelength. Both the critical
wavelength and the faceted island widths scale as $\varepsilon^{-2}$,
as observed in experiments \cite{floro,dorsch,pidduck,koo} in which
islands develop from long ripple like structures (corresponding to our
model of plane strain).

All results discussed henceforth refer to Ge/Si(001) though the same
trends were seen in Ge$_{0.5}$Si$_{0.5}$/Si(001). Islands form in a
'chain-reaction ripple' effect (i.e., islands tend to develop near
other islands) as is illustrated in Fig.\ \ref{facevolc25}. This mode
of growth has also been observed in experiment \cite{tromp,sutter}. The
ripple effect occurs because the growth of the island destabilizes the
flat film at its boundaries. After initial island formation we observe
island ripening occurring over much longer time scales (about 50 times
longer).

During annealing the islands are fully faceted. Their tops are faceted
at $0^{\circ}$ and their sides at $45^{\circ}$. This shape is preserved
as the islands grow, i.e., the islands maintain a fixed diameter-height
ratio (as seen in experiment \cite{floro,moison,kamins,mo} and theory
\cite{long}). During deposition, on the other hand, an interesting
transition is observed in the island shape. Initially, the islands are
fully faceted as during annealing. However, when the islands reach a
certain diameter, they stop growing laterally and only vertical growth
occurs. This critical diameter is about 40nm for Ge islands grown on a
Si(001) substrate (during annealing we never observed islands which
exceeded this diameter). Thus the islands become tall and narrow, and
their sides are steeper than $45^{\circ}$. This shape transition is
observed experimentally \cite{ross,kamins,rastelli,vostokov}, and is
sometimes referred to as the ``pyramid-to-dome transition''. The
driving force behind it is the increased elastic relaxation experienced
by tall-narrow islands. Theoretical equilibrium calculations with
isotropic surface tension \cite{spencertersoff} show a continuous
increase in island aspect ratio with increasing island volume, as
elastic effects dominate surface tension effects. The sharp rather than
smooth transition in growth mode we observe is due to the anisotropic
nature of the surface tension and in particular the presence of a facet
at $45^\circ$. Note that contrary to existing explanations of the
island shape transition (see e.g.\ \cite{darukatersoff}), the
transition occurs without an additional facet orientation at a larger
angle.

\begin{figure}[h]
 \centerline{
   \epsfxsize=90mm
   \epsffile{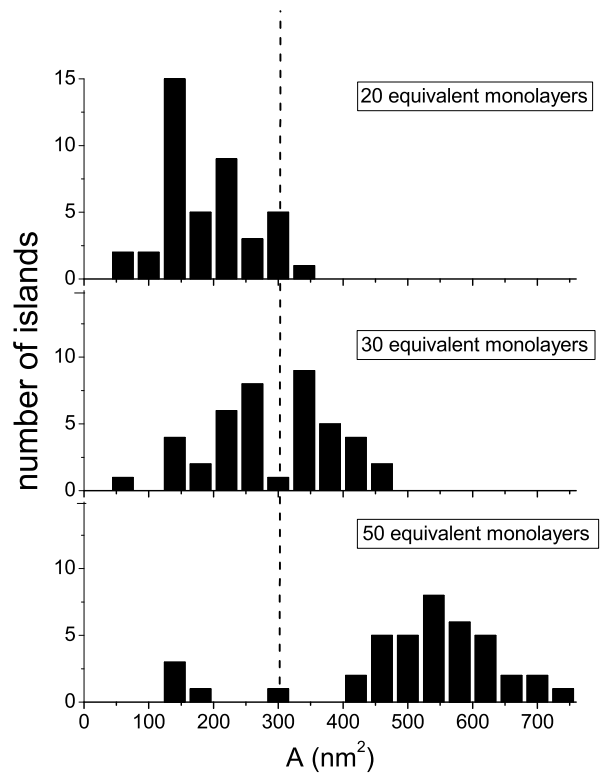}}
    \caption{
Distribution of island cross-sectional area, A (recall that islands are
infinitely long in the $z$-direction), during directed deposition of Ge
on a Si(001) substrate. The rate of deposition is 5.2nm/s, and the
initial film height is 10 monolayers. The dashed vertical line shows
the separation between the early growth mode in which the island
height-width ratio is preserved and the later vertical growth mode. }
     \label{depstat}
\end{figure}

The transition in island shape and growth mode is clearly reflected in
the size distribution shown in Fig.\ \ref{depstat}. Narrow island size
and spacing distributions are seen during early deposition (see Fig.\
\ref{depstat}, 20 equivalent monolayers). These narrow distributions
are observed in many experiments
\cite{ross,floro,kamins,mo,vostokov,medeiros,floro2,kastner}. During
later deposition (30 equivalent monolayers) a {\em bimodal}
distribution forms as some of the islands pass from the fully faceted
to the tall-narrow shape. At later times (e.g.\ 50 equivalent
monolayers) nearly all islands have the tall-narrow shape. At this
stage the distribution becomes quite symmetric and evolves at a fixed
distribution width (increasing its mean). Similar results were observed
in experiment \cite{ross,kamins,rastelli,vostokov}.

\begin{figure}[h]
   \centerline{
   \epsfxsize=90mm
   \epsffile{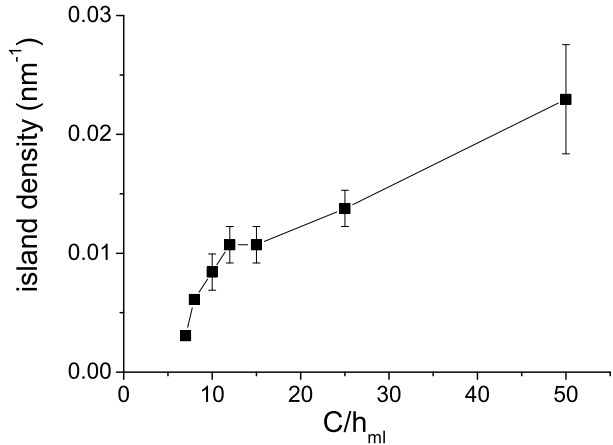}}
\caption{Island density in a stable array of Ge islands on a Si(001)
substrate after ripening has ended. $C$ is the initial flat film
thickness, and $h_{ml}$ is the thickness of one monolayer. The error
bars refer to the span of island densities observed with different
initial surface morphologies. }
     \label{idensity}
\end{figure}

One of our central observations is that annealing of a perturbed flat
film with anisotropic surface tension leads to the formation of a
stable array of islands. This result is consistent with several
experimental systems \cite{medeiros,ozkan,cirlin,miller}, and is in
contrast with films of isotropic surface tension where the islands
ripen indefinitely. Theoretical studies also predict stable island
arrays \cite{shchukin,daruka,chiu,zhang}. The crucial term in
determining the stability of an island array apart from anisotropic
surface tension and a film-substrate interaction is the presence of an
elastic contribution due to island edges. This contribution is
automatically present in our calculations and does not need to be
introduced separately. Theoretical works that ignore this term
\cite{floro2,terlegoues} predict continuous ripening.

Our simulations show that the density of islands in the stable array
increases with increasing film thickness (see Fig.\ \ref{idensity}). An
increase in island density with film thickness has also been seen in
many experiments
\cite{mo,kastner,miller,leonard,ramachandran,kobayashi,kamins2}. Indeed
Miller et al.\ \cite{miller} and Kamins et al.\ \cite{kamins2}
performed annealing experiments and Leonard et al.\ \cite{leonard}
performed experiments with very small deposition rates. These three
experiments clearly show the increase in island density as film
thickness increases. This result was predicted by Daruka and Barabasi
\cite{daruka} in minimal energy equilibrium calculations. Here we show
for the first time that the increase in island density also results
from evolution simulations. This observation is particularly important,
since other evolution studies \cite{chiu} predicted a decrease in
island density with increasing film thickness. We believe this is due
to the smooth form of surface tension used in \cite{chiu}. Indeed when
we carried out simulations with a smooth form of surface tension
similar to that used in \cite{chiu}, we also observed a decrease in
island density. This clearly demonstrates the importance of using a
cusped form of surface tension to accurately model evolution of
faceting surfaces.

\begin{figure}[h]
   \centerline{
   \epsfxsize=90mm
   \epsffile{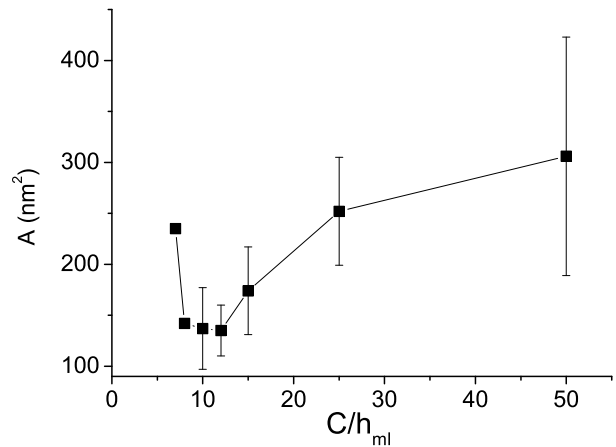}}
\caption{Average island cross-sectional area, A (recall that islands
are infinitely long in the $z$-direction), in a stable array of Ge
islands on a Si(001) substrate after ripening has ended. $C$ is the
initial flat film thickness, and $h_{ml}$ is the thickness of one
monolayer. The error bars refer to the standard deviation in island
sizes observed throughout all samples of the same film thickness.}
     \label{isize}
\end{figure}

As can be seen in Fig.\ \ref{isize}, the island size also shows a
slight increase with increasing film thickness, with islands increasing
in width from 25nm to 40nm, and cross-sectional area from $100nm^{2}$
to $500nm^{2}$ (recall that islands are infinitely long in the
$z$-direction). The island size at 7 monolayers is larger than expected
due to finite size effects. Note that even the smallest islands have a
finite non-zero size. Experiments indeed see islands forming only above
a certain size which increases with increasing film thickness
\cite{mo,medeiros,floro2,kastner,kamins2}. However, as the experiments
which were performed for annealing and showed stable island arrays
tended not to vary the film thickness, it is difficult to compare our
results with experimental observations. Our result is in accordance
with that predicted in equilibrium calculations by Daruka and Barabasi
\cite{daruka}.

When surface tension is isotropic, corresponding to films above the
roughening transition temperature, flat film evolution during annealing
is very different from that described above. Perturbations in films
thinner than $h_c$ decay, and flat films with thickness
$h_c<h<h_{c}+\Delta$, where $\Delta\approx 1$ monolayer, develop a
stable smooth wavy morphology at $\lambda_{c}$. That is, perturbations
of other wavelengths decay and perturbations of wavelength
$\lambda_{c}$ grow to a finite amplitude. This is a mode of growth
neither seen nor predicted before. Stable, non flat morphology, has
previously only been predicted for faceting films
\cite{shchukin,daruka,chiu,zhang}. In fact, other groups maintain that
isotropic films should be unstable to ripening
\cite{chiu,zhang,kukta,spencer}. While films are linearly unstable to
perturbations of wavelengths $\lambda_{-}\leq \lambda \leq
\lambda_{+}$, our simulations show that the nonlinearity stabilizes the
growth of wavelengths close to $\lambda_{-}$ and $\lambda_{+}$. As a
result the growth of the perturbation saturates and stops at a finite
amplitude, as seen by Spencer and Meiron \cite{spencermeiron} for
infinitely thick films. When films are sufficiently close to the linear
wetting layer thickness, the range of nonlinear saturation extends over
the entire range of linearly unstable wavelengths and so a stable wavy
morphology is observed. For films thicker than $h_{c}+\Delta$,
initially a wavy structure at the most unstable wavelength forms. The
hills of these waves then ripen on larger and larger length scales,
until isolated islands are left that continue ripening.

This work was supported by the Israeli Science Foundation.


\begin{thebibliography}{10}

\bibitem[*]{email1}
hrg1000@wicc.weizmann.ac.il.

\bibitem[**]{email2}
daniel.kandel@weizmann.ac.il, http://www.weizmann.ac.il/$\sim$fekandel.

\bibitem{ross}
F.M. Ross, J. Tersoff, R.M. Tromp, Phys. Rev. Lett. \textbf{80}, 984
(1998).

\bibitem{mullins}
W.W. Mullins, J. Appl. Phys. {\bf 28}, 333 (1957).

\bibitem{eisenberg1}
H.R Eisenberg, D. Kandel, Phys. Rev. Lett. \textbf{85}, 1286 (2000).

\bibitem{eisenberg2}
H.R Eisenberg, D. Kandel, Phys. Rev. B {\bf 66}, 155429 (2002).

\bibitem{bonzel}
H.P. Bonzel, E. Preuss, Surf. Sci. {\bf 336}, 209 (1995).

\bibitem{eisenberg3}
H.R Eisenberg, D. Kandel, E. Kaxiras, I.N. Remediakis, in preparation.

\bibitem{floro}
Floro J.A, Chason E, Twesten R.D, R.Q. Hwang, L.B. Freud, Phys. Rev.
Lett. {\bf
  79}, 3946 (1997); J.A. Floro, E. Chason, L.B Freund, R.D Twesten, R.Q. Hwang,
  G.A Lucadamo, Phys. Rev. B \textbf{59}, 1990 (1999).

\bibitem{dorsch}
W.Dorsch, B. Steiner, M. Albrecht, H.P Strunk, H. Wawra, G. Wagner, J.
Cryst.
  Growth \textbf{183}, 305 (1998).

\bibitem{pidduck}
A.J. Pidduck, D.J. Robbins, A.G. Cullis, W.Y. Leong, A.M. Pitt, Thin
Solid
  Films, \textbf{222}, 78 (1992).

\bibitem{koo}
B.H. Koo, T. Hanada, H. Makino, T. Yao, Appl. Phys. Lett. \textbf{79},
4331
  (2001).

\bibitem{tromp}
R.M Tromp, F.M. Ross, M.C Reuter, Phys. Rev. Lett. \textbf{84}, 4641
(2000).

\bibitem{sutter}
P. Sutter, M.G. Lagally, Phys. Rev. Lett. \textbf{84}, 4637 (2000).

\bibitem{moison}
J.M. Moison, F. Houzay, L. Leprince, E. Andre, O. Vatel, Appl. Phys.
Lett.
  \textbf{64}, 196 (1994).

\bibitem{kamins}
Kamins T.I, Carr E.C, Williams R.S, Rosner S.J, J. Appl. Phys.
\textbf{81}, 211
  (1997).

\bibitem{mo}
Mo Y.-W, Savage D.E, Swartzentruber B.S, Lagally M.G, Phys. Rev. Lett.
  \textbf{65}, 1021 (1990).

\bibitem{long}
F. Long, S.P.A. Gill, A.C.F. Cocks, Phys. Rev. B, \textbf{64}, 121307
(2001).

\bibitem{rastelli}
A. Rastelli, H. von K\"{a}nel, Surf. Sci. Lett. \textbf{515}, L493
(2002).

\bibitem{vostokov}
N.V. Vostokov et al. J. Cryst. Growth, \textbf{209}, 302 (2000).

\bibitem{spencertersoff}
B.J. Spencer, J. Tersoff, Phys. Rev. Lett. \textbf{79}, 4858 (1997).

\bibitem{darukatersoff}
I. Daruka, J. Tersoff, A.-L Barab\'{a}si, Phys. Rev. Lett. \textbf{82},
2753
  (1999).

\bibitem{medeiros}
G. Medeiros-Ribeiro, T.I. Kamins, D.A.A. Ohlberg, R. Stanley Williams,
Phys.
  Rev. B, \textbf{58},3533 (1998).

\bibitem{floro2}
J.A. Floro, G.A. Lucadamo, E. Chason, L.B. Freund, M. Sinclair, R.D.
Twesten,
  R.Q. Hwang, Phys. Rev. Lett. \textbf{80}, 4717 (1998); J.A. Floro, M.B.
  Sinclair, E. Chason, L.B. Freund, R.D. Twesten, R.Q. Hwang, G.A. Lucadamo,
  Phys. Rev. Lett. \textbf{84}, 701 (2000).

\bibitem{kastner}
M. Kastner, B. Voigtlander, Phys. Rev. Lett. \textbf{82}, 2745 (1999).

\bibitem{ozkan}
C.S. Ozkan, W.D. Nix, H.Gao, Appl. Phys. Lett. \textbf{70}, 2247
(1997).

\bibitem{cirlin}
G.E. Cirlin, G.M. Guryanov, A.O. Golubok, S. Ya. Tipissev, N.N.
Ledentsov, P.S.
  Kop'ev, M. Grundmann, D. Bimberg, Appl. Phys. Lett. \textbf{67}, 97 (1995).

\bibitem{miller}
M.S. Miller, S. Jeppesen, D. Hessman, B. Kowalski, I. Maximov, B.
Junno, L.
  Samuelson, Solid State Electon. \textbf{40}, 609 (1996).

\bibitem{shchukin}
V.A. Shchukin, N.N. Ledentsov, P.S. Kop'ev, D.Bimberg, Phys. Rev. Lett.
  \textbf{75}, 2968 (995).

\bibitem{daruka}
I. Daruka and A.-L. Barab\'{a}si, Phys. Rev. Lett. {\bf 79}, 3708
(1997).

\bibitem{chiu}
C.-h. Chiu, Appl. Phys. Lett. \textbf{75}, 3473 (1999).

\bibitem{zhang}
Y.W. Zhang, Phys. Rev.B \textbf{61}, 10388 (2000).

\bibitem{terlegoues}
J. Tersoff, F.K. LeGoues, Phys. Rev. Lett. \textbf{72}, 3570 (1994).

\bibitem{leonard}
D. Leonard, K. Pond, P.M. Petroff, Phys. Rev. B, \textbf{50}, 11687
(1994).

\bibitem{ramachandran}
T.R. Ramachandran, R. Heitz, P. Chen, A. Madhukar, Appl. Phys. Lett.
  \textbf{70}, 640 (1997).

\bibitem{kobayashi}
N.P. Kobayashi, T.R. Ramachandran, P. Chen, A. Madhukar, Appl. Phys.
Lett.
  \textbf{68}, 3299 (1996).

\bibitem{kamins2}
T.I. Kamins, G. Medeiros-Ribeiro, D.A.A. Ohlberg, R. Stanley Williams,
J. Appl.
  Phys. \textbf{85}, 1159 (1999).

\bibitem{kukta}
R.V. Kukta, L.B. Freund, J. Mech. Phys. Solids. {\bf 45}, 1835 (1997).

\bibitem{spencer}
B.J. Spencer, J. Tersoff, Phys. Rev. Lett. \textbf{79}, 4858 (1999).

\bibitem{spencermeiron}
B.J. Spencer, D.I. Meiron, Acta. Metall. Mater. {\bf 42}, 3629 (1994).

\end{thebibliography}
\end{document}